\documentclass[preprint2]{aastex631}

\usepackage{natbib}

\submitjournal{ApJ}
\received{}
\revised{}
\accepted{}

\shorttitle{Starspot temperature of CoRoT-2 from multiwavelength}
\shortauthors{Valio et al.}

\begin{document} 

\title{Starspot temperature of CoRoT-2 from multiwavelength observations with SPARC4}

%\titlerunning{Starspot temperature}

%   \subtitle{Starspot temperature}

\correspondingauthor{Adriana Valio}
\email{avalio@craam.mackenzie.br}

\author[0000-0002-1671-8370]{Adriana Valio}
\affiliation{Centro de Rádio Astronomia e Astrofísica Mackenzie, Universidade Presbiteriana Mackenzie, Rua da Consolação, 930, SP, Brazil}

\author[0000-0002-5084-168X]{Eder Martioli}
\affiliation{Laboratório Nacional de Astrofísica, Rua Estados Unidos, 154, Itajubá, MG, Brazil}

\author[0009-0004-9308-2072]{Andre O. Kovacs}
\affiliation{Centro de Rádio Astronomia e Astrofísica Mackenzie, Universidade Presbiteriana Mackenzie, Rua da Consolação, 930, SP, Brazil}

\author[0000-0002-7375-8088]{Viktor Y. D. Sumida}
\affiliation{Centro de Rádio Astronomia e Astrofísica Mackenzie, Universidade Presbiteriana Mackenzie, Rua da Consolação, 930, SP, Brazil}

\author[0000-0001-8179-1147]{Leandro de Almeida}
\author{Diego Lorenzo-Oliveira}
\affiliation{Laboratório Nacional de Astrofísica, Rua Estados Unidos, 154, Itajubá, MG, Brazil}
\affiliation{SOAR Telescope/NSF’s NOIRLab, Avda Juan Cisternas 1500, 1700000, La Serena, Chile
}

\author{Francisco Jablonski}
\author[0000-0002-9459-043X]{Claudia V. Rodrigues}
\affiliation{Instituto Nacional de Pesquisas Espaciais/MCTI, Av. dos Astronautas, 1758, São José dos Campos, SP, Brazil }

\begin{abstract} 
Measuring starspot temperatures is crucial for understanding stellar magnetic activity, as it affects stellar brightness variations, influences exoplanet transit measurements, and provides constraints on the physical conditions and energy transport in active regions, offering insights into stellar dynamos.
Our goal is to determine the temperature of starspots on the active star CoRoT-2  to enhance our understanding of magnetic activity in young, solar-like stars.
Multiwavelength observations were conducted using the SPARC4 instrument on the 1.6-m telescope at Pico dos Dias Observatory (Brazil), capturing simultaneous transit data in four photometric bands (g, r, i, and z). The \texttt{ECLIPSE} model, combined with MCMC fitting, was used to model spot characteristics during the planetary transit of CoRoT-2 b.  The spot intensities were analyzed considering three different methods:  the assumption of blackbody emission, the PHOENIX atmospheric model, and multiwavelength fitting assuming the same spot parameters for all wavelengths.
Two starspots were detected in the residuals of the light curve, yielding temperature estimates of 5040 -- 5280 K based on the three different methods. These values align more closely with the temperatures of solar penumbrae than with typical umbral temperatures, suggesting relatively moderate magnetic activity.  The radius of the spots ranged from 0.34 -- 0.61 the planetary radius, or equivalently (38 -- 69)$\times10^6$m, much larger than sunspots.
This study provides a method to estimate spot temperatures on active stars using multiband photometry, with results indicating penumbral-like temperatures on CoRoT-2. The methodology enhances precision in starspot temperature estimation, beneficial for studies of stellar activity and exoplanet characterization.

\end{abstract}

%
%-------------------------------------------------------------------

\section{Introduction}

Starspots are regions of intense magnetic activity on a star’s surface. Estimating their temperature helps to better understand the star’s magnetic field dynamics, including the mechanisms driving stellar activity and variability. Moreover, starspots can affect the precision of exoplanet parameter estimates based on transit photometry and transmission spectroscopy. When an exoplanet passes in front of a star, the presence of cool/dark starspots can distort the measured light curve, leading to inaccuracies in the estimation of the planet’s size, orbital parameters, and atmosphere \citep{Tregloan2019}. These distortions can be corrected by accurately estimating the temperature of starspots, thus improving the accuracy of exoplanet characterization.

Similar to sunspots on the Sun \citep{Solanki2003}, the temperature of starspots offers valuable insights into a star’s age and magnetic activity cycles. Cooler spots are associated with stronger magnetic fields, which are directly linked to the star’s magnetic activity and provide important information about its dynamo processes and evolutionary state \citep{berdy+05}.
This magnetic activity, traced by the presence and properties of starspots, also has significant implications for the habitability of surrounding exoplanets \citep{Airapetian2020}. While cooler starspots can locally reduce stellar radiation, increased magnetic activity often leads to more intense space weather events, which may erode planetary atmospheres and pose challenges to the development or sustainability of life.

CoRoT-2 is a G7V-type star, with an apparent V magnitude of 12.57 , located approximately 930 light-years from Earth. It has an estimated mass of about 0.972 M$_{\odot}$ and a radius of 0.938 R$_{\odot}$ \citep{Stassun2019}. The star’s effective temperature ($T_{\text{eff}}$) is $5529\pm110$ K \citep{Stassun2019}, making it slightly cooler than the Sun. \cite{Guillot2011} set its maximum age at 500 Myr, which is estimated at $\sim$128 million years by \cite{Hamer2022}.   

Its close-in transiting gas-giant exoplanet, CoRoT-2 b, orbits at a distance of just 0.028 au from the star and completes an orbit in 1.743 days \citep{alonso2008transiting}. CoRoT-2 b has a mass of approximately 3.31 M$_{\text{J}}$ and a radius of 1.465 R$_{\text{J}}$ \citep{Bruno2016}, indicating that it is significantly inflated, likely due to the intense irradiation from its host star. 

A highly active star, CoRoT-2 exhibits extensive spot coverage, which significantly impacts the observed light curves during planetary transits.
The many spots on its surface have been mapped using different techniques \citep{lanza2009, Wolter2009, Huber2009, Czesla2009}, including transit mapping by \cite{silva2010}. 

By analyzing 77 consecutive transits observed by the CoRoT satellite \citep{Baglin2006}, \cite{silva2010} identified 369 starspots. The spots range in size from 0.2 to 0.7 times the planet's radius ($R_{plan}$), with an average size of 0.46 $R_{plan}$, translating to approximately 100,000 km in diameter. Spot intensities varied between 30\% and 80\% of the photospheric intensity, indicating a mean spot temperature of 4700 $\pm$ 300 K, about 800 K cooler than the stellar photosphere (5529 K).  These characteristics suggest that CoRoT-2’s spots are considerably larger and cooler than typical sunspots, reinforcing the star's classification as more magnetically active than the Sun.

\cite{Schutte2023} analyzed the characteristics of starspots on HAT-P-11, a K4 star ($T_{eff} = 4780$ K) with an estimated age of 6.5 Gyr \citep{Bakos2010}, using high-precision multifilter photometry to measure the starspot temperature. Transits of HAT-P-11 b were captured through MuSCAT3 at Haleakala Observatory with filters, including SDSS g, r, i, and Z, which maximize spot contrast differences, especially when using SDSS g and i filters. 
However, the authors were not able to measure the temperature directly, since there was only spot signal in one of the bands. Nevertheless, they modeled what the signal would be for an average starspot temperature of 4500 $\pm$ 100 K.

As reviewed by \citet{berdy+05}, spot temperature contrasts vary widely across stellar types and observational techniques, but consistently highlight the importance of including cool surface features in transit light-curve and spectral modeling. Recent studies have emphasized the impact of stellar heterogeneities, particularly starspots, on the interpretation of transmission spectra, as these features can significantly bias the retrieval of atmospheric properties. For instance, \citet{Fournier-Tondreau2024} reanalyzed JWST/NIRISS observations of the exoplanet HAT-P-18 b and showed that unocculted starspots on the K dwarf host can alter molecular abundance retrievals, emphasizing the need for joint modeling of planetary and stellar signals. The authors also determined a small spot–photosphere temperature contrast of $\Delta T = -93 \pm 15$ K, corresponding to an occulted spot temperature of 4710 $\pm 15$ K. On the other hand, \citet{Libby-Roberts2023} found larger contrasts in the mid-M dwarf TOI-3884, with occulted spot temperatures between 2700 and 2900 K ($\Delta T$ = -480 to -280 K) based on photometry and radial velocities, while \citet{Almenara2022}, using multiband near-IR transit photometry, reported a temperature difference of $\Delta T = -187 \pm 21$ K for the same star.

Here we report on the simultaneous multiwavelength observations of CoRoT-2 with the new SPARC4 multiband imager and polarimeter on the 1.6-m Perkin Elmer telescope at Pico dos Dias Observatory (OPD) in Brazil. The observations are detailed in Section~\ref{sec:obs}, while the spot modeling is described in Section~\ref{sec:model}. The results of the spots' temperature are reported in Section~\ref{sec:temp}, and the final section presents the discussion and main conclusions.

%--------------------------------------------------------------------
\section{Observations of CoRoT-2 with SPARC4}\label{sec:obs}

%\begin{figure}
\begin{figure*}
    \centering
\includegraphics[width=0.45\linewidth]{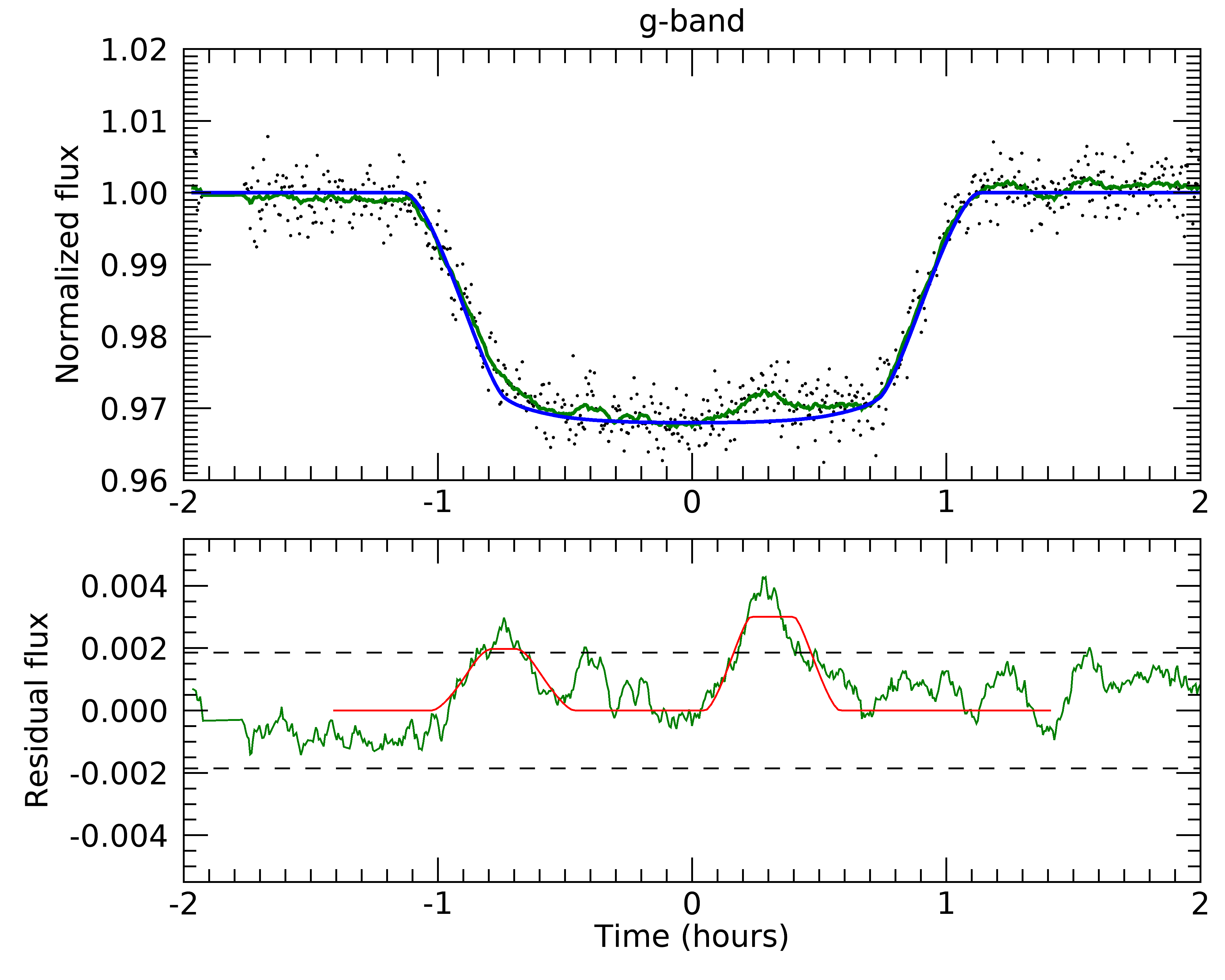}
\includegraphics[width=0.45\linewidth]{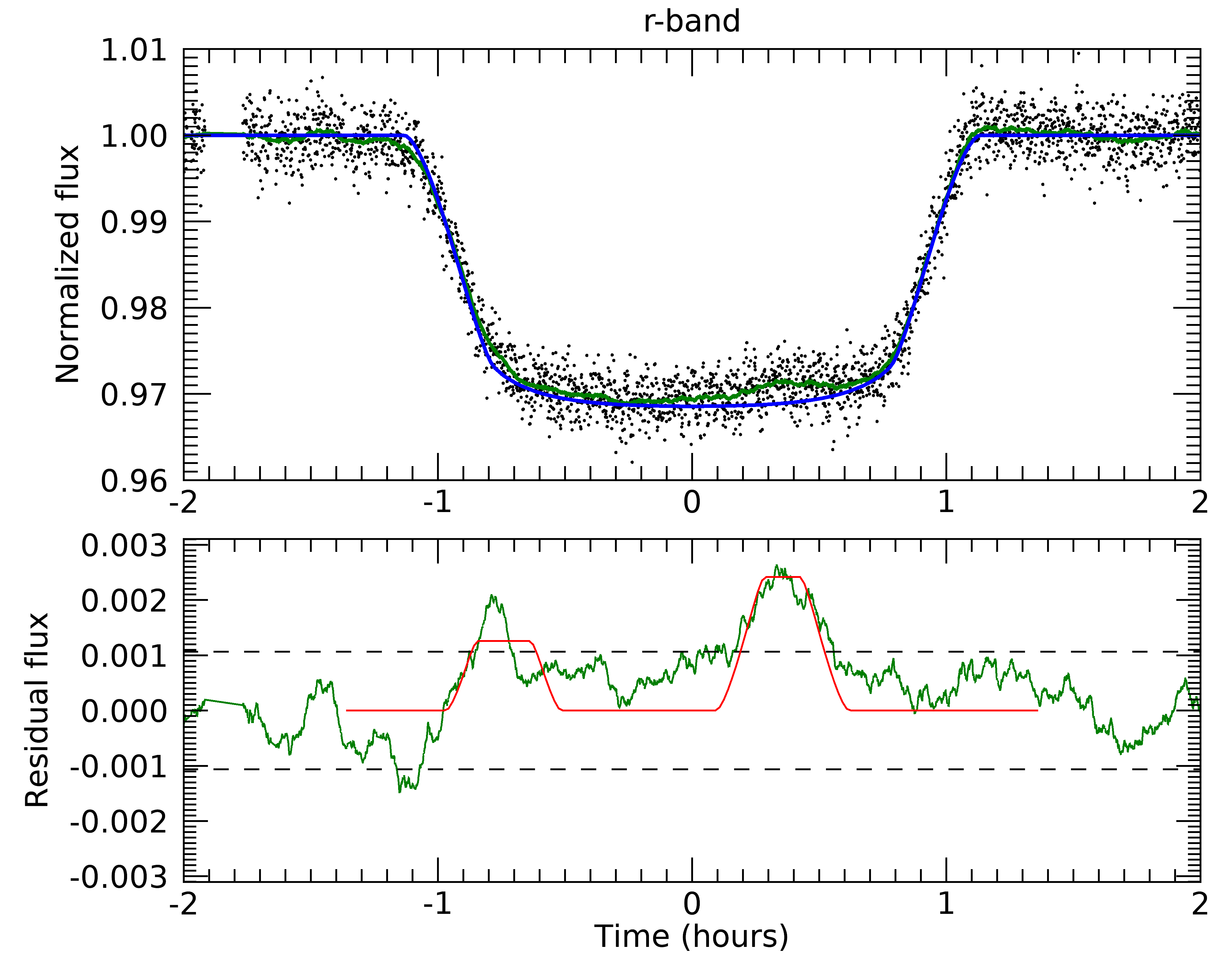}
\includegraphics[width=0.45\linewidth]{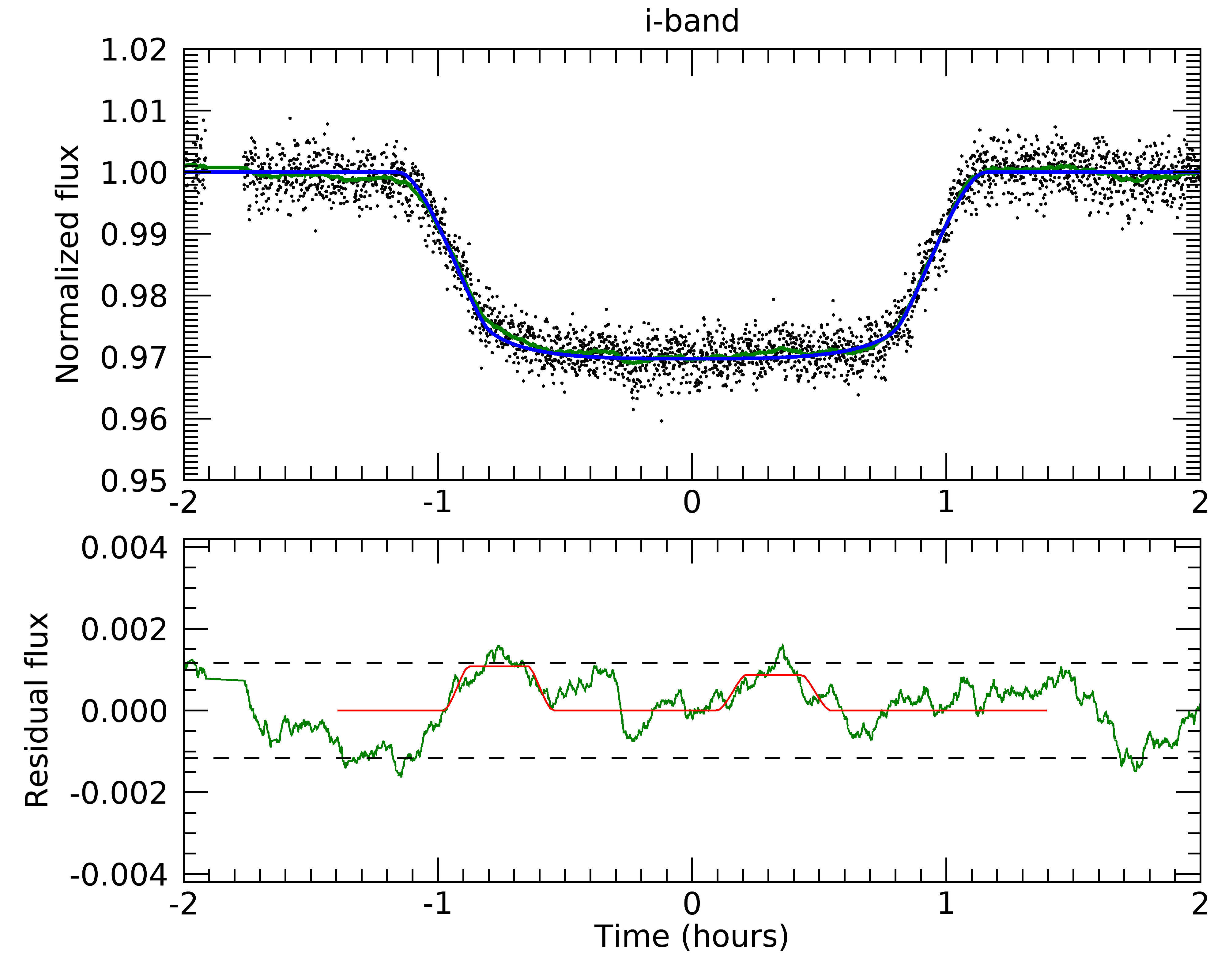}
\includegraphics[width=0.45\linewidth]{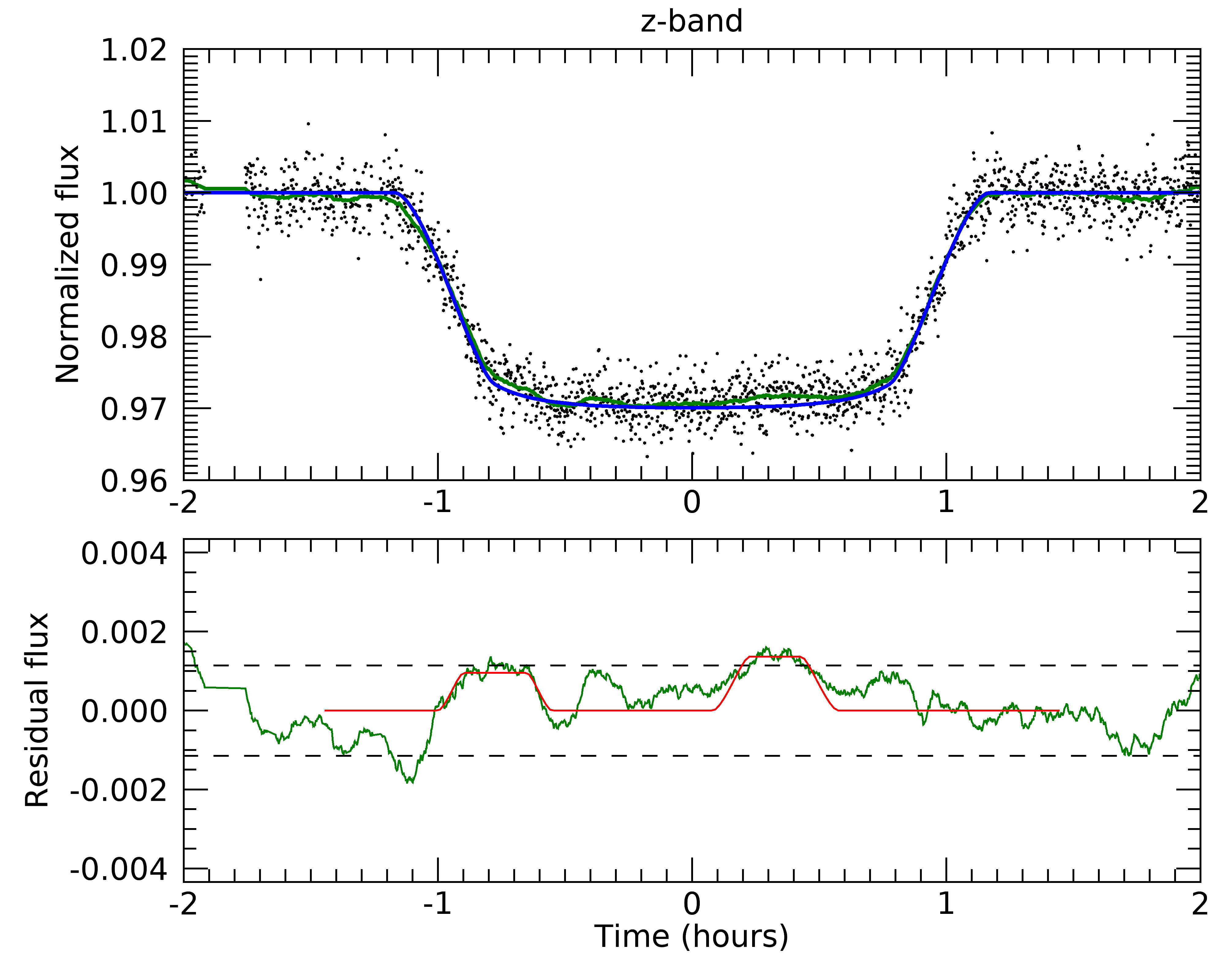}
    \caption{Light curves of the transit of CoRoT-2 b in the four photometric bands of SPARC4, with the smoothed data shown in green and a model of a spotless star depicted in blue. The bottom panels of each waveband shows the residuals, after subtraction of a spotless star model (blue curve in the top panel). The time at transit center is 2,460,507.10 BJD.
    The red curves show the result of the model fitting for two spots. The horizontal dashed lines represent a 2$\sigma$ threshold for the signal to be considered due to a spot ($1\sigma = 930, 530, 580,$ and $570$ ppm for the smoothed signal from bands g, r, i, and z, respectively).}
    \label{fig:transits}
%\end{figure}
\end{figure*}

Two transits of the young solar-like star CoRoT-2 were observed under program OP2024A-004 (PI: E. Martioli) with the SPARC4 instrument installed on the 1.6-m telescope at OPD: the first transit on June 20th and the second on July 4th, 2024. Unfortunately, the second transit cannot be used due to severe cloud cover during the observations.

The SPARC4 (Simultaneous Polarimeter and Rapid Camera in Four bands) is an advanced instrument designed for use on the 1.6-m telescope at the \textit{Observatório do Pico dos Dias} (OPD) in Brazil \citep{Rodrigues2012, Rodrigues2024}. It is a multichannel imager that allows for simultaneous observations in four photometric bands. 
The SPARC4 instrument was operated in photometric mode, with the four CCDs configured in Conventional mode, a preamp value of `Gain 2', and a readout rate of 1~MHz, with frame transfer enabled. Individual exposure times per frame were set to 20~s, 5~s, 5~s, and 7.5~s in the g, r, i, and z bands, respectively. The overhead between frames for the adopted modes is negligible \citep[see ][]{Bernardes2024}.

The data were reduced using the SPARC4 Pipeline\footnote{\url{https://github.com/edermartioli/sparc4-pipeline}} (Martioli et al., in prep.), which performs bias subtraction followed by flat-field and gain corrections. The pipeline uses the packages ASTROPOP\footnote{\url{https://github.com/juliotux/astropop/tree/main/astropop}} \citep{Campagnolo2019} and \texttt{Astropy}\footnote{\url{https://www.astropy.org/}} \citep{Astropy2013} for data processing. The pipeline detects sources in a stack of a few dozen well-exposed frames and registers all individual frames via cross-correlation with the stack. Aperture photometry is performed on each source using the package \texttt{Photutils} \footnote{\url{https://photutils.readthedocs.io/en/stable/}} \citep{Photutils2016} with several aperture radii, ranging from 5 to 25 pixels.
The best photometric results were found for an aperture radius of 12 pixels, corresponding to approximately 4 arcseconds, which was utilized in all subsequent steps.
Finally, the pipeline performs differential photometry using the flux sum of multiple sources as comparison, selecting only sources with magnitudes similar to CoRoT-2 and discarding bright, saturated sources. The resulting light curves are plotted in the top panels of Figure~\ref{fig:transits} for the four bands.
The points are the observations and the green curve is the smoothed flux with a running mean every 10 points for the g-band and 40 points for all others.

\section{Spot modeling}\label{sec:model}

The \texttt{ECLIPSE}\footnote{\url{https://github.com/Transit-Model-CRAAM/pipelineMCMC}} code, initially developed by \cite{silva+03}, is a computational tool designed to simulate the light curves of star-planet systems where the planet transits across the stellar disk. \texttt{ECLIPSE} specifically models the detailed effects of stellar spots and faculae on transit light curves, including limb darkening and spot foreshortening. 
This spot model has been successfully applied to G-type \citep{silva2010,zaleski+19,netto2020, valio2024}, K-type \citep{araujo+21,valio22,zaleski2022dynamo}, and M-type  \citep{zaleski2020} stars.

Before modeling the spots, it is necessary to first model the planetary transit in order to determine the planetary radius, orbital parameters, and limb darkening coefficients. This is done by applying the \texttt{ECLIPSE} model for a star without spots.
The limb darkening coefficients are  free parameters of the model and are not the same at each wavelength.
The quadratic law for limb darkening for the intensity, $I(\mu)$, at a point on the stellar disk, is given by:
\begin{equation}
    \frac{I(\mu)}{I(1)} = 1 - u_1 (1 - \mu) - u_2 (1 - \mu)^2
    \label{eq:limb}
\end{equation}
\noindent where $\mu = cos(\theta)$ is the cosine of the angle between the line-of-sight and the normal to the stellar surface,
$I(1)$ is the intensity at the center of the disk, and $u_1$ and $u_2$  are the limb darkening coefficients.

The stellar ($u_1$ and $u_2$) and planetary parameters at each wavelength were derived through MCMC fits to the transit data and are represented by the blue curve in the top panels of Figure~\ref{fig:transits}.
In these fits, certain parameters were held fixed, including the orbital period (1.7429923 $\pm$ 0.0000024 day), semi-major axis (65.498 $\pm$ 0.029 $R_{star}$), and inclination angle (87.14$^\circ$ $\pm$  0.18$^\circ$) which were taken from \cite{alonso2008transiting}. 
The resulting parameters from the MCMC analysis are listed in Table~\ref{tab:param} and the fit is shown as the blue curve in the top panels of Figure~\ref{fig:transits}. 

\begin{table}[]
    \centering
    \begin{tabular}{c|cc|c}
\hline
\hline
band & $u_1$ & $u_2$ &  $R_{plan}$ ($R_{star}$)    \\
\hline
g & 0.0003 $\pm$ 0.022 & 0.46 $\pm$ 0.11 & 0.1686 $\pm$ 0.0015  \\
r & 0.070 $\pm$ 0.015 & 0.47 $\pm$ 0.04 & 0.1654 $\pm$ 0.0003\\
i & 0.001 $\pm$ 0.018 &  0.60 $\pm$ 0.03 & 0.1641 $\pm$ 0.0003 \\
z & 0003 $\pm$ 0.011 & 0.468 $\pm$ 0.024 & 0.1645 $\pm$ 0.0003  \\
\hline
    \end{tabular}
    \caption{Stellar and planetary parameters obtained from the MCMC fit to the transit light curves in four photometric bands (g, r, i, z). The table lists the limb-darkening coefficients ($u_1$ and $u_2$), and the planetary radius in units of stellar radius ($R_{plan}/R_{star}$). The orbital parameters such as period, semi-major axis, and inclination angle were fixed during the fitting process.}
    \label{tab:param}
\end{table}

Next, each spot is modeled by its radius (with respect to the planetary radius), intensity (relative to the maximum disk center intensity), and position (latitude and longitude). 
To better identify a spot signal in the transit light curve, we first apply a running mean to smooth the light curve and then subtract the model of a spotless star (blue curve) from each light curve. The result is the residual, which is plotted as a green curve in the lower panels of Figure~\ref{fig:transits}.

Here we only consider as spot signatures the features ("bumps") that are present in the residuals of all 4 bands and that exceed the 2$\sigma$ threshold (horizontal dashed lines in the bottom panels of Figure~\ref{fig:transits}). According to these criteria,
two spot signatures, or "bumps", were identified in the residual light curve: one at 0.3 h and the other at -0.8 h from transit center, these spot crossing times correspond to approximate stellar longitudes of $20^\circ$ and $-50^\circ$, respectively.

We modeled both spot signatures at each wavelength using the \texttt{ECLIPSE} code with MCMC. 
The latitude of both spots was considered a fixed parameter and assumed to be located at the center of the transit latitude band: 
\begin{equation}
    lat_{spot} = -\arcsin{[a\cdot\cos(i)]}
    \label{eq:lat}
\end{equation}
\noindent where $a$ is the orbital semi-major axis (in units of stellar radius) and $i$ its inclination angle of the orbit. The resulting latitude of -23.7$^\circ$ was 
 arbitrarily chosen to be located in the Southern Hemisphere, hence the minus sign.

 In fact, the program \texttt{ECLIPSE} allows for the latitude to have any value, not necessarily the center of the transit chord, as given by Equation~\ref{eq:lat}. However, varying the latitude of the spots would introduce an additional free parameter in the fitting procedure, increasing the model's complexity and potential degeneracies. Indeed, spots could be larger and offset from the center of the transit chord, resulting in grazing occultations. In such cases, the planet may not cross the umbral core of the spot, but instead only its penumbral region. This would naturally lead to higher inferred spot temperatures, more consistent with penumbral values observed in solar spots.

The MCMC results of the spot parameters: radius (in units of planetary radius), intensity (relative to stellar central intensity), and longitude are listed in Table~\ref{tab:spot} and the fit is plotted as the red curve in the bottom panels of Figure~\ref{fig:transits}. 
In this analysis, the prior distributions adopted for the MCMC sampling are uniform (flat) priors with hard boundaries. Specifically, the prior function imposes the following constraints on the parameters: the spot radius is restricted to the interval 0.1 to 1.5 (in units of planetary radii), the spot intensity must lie between 0.1 and 0.95 (relative to the stellar disk center intensity), and the spot longitude is limited to the range between $10^\circ$ and $30^\circ$ for spot 1 and $-60^\circ$ and $-40^\circ$ for spot 2. 
The MCMC samples and posterior distributions are illustrated in the corner plots of Figures~\ref{fig:spot1} and \ref{fig:spot2}.

In Section~\ref{sec:multi}, we model each spot with MCMC but considering the same radius, longitude, and temperature for all four bands. The latitude remains fixed at -23.7$^\circ$.

\begin{table}[]
    \centering
    \begin{tabular}{c|c|ccc}
\hline
\hline
& Band & R$_{spot}$ (R$_{plan}$) & Intensity & longitude ($^\circ$)  \\
\hline
Spot 1 & g &  0.53$^{+0.19}_{-0.09}$ & 0.61$^{+0.17}_{-0.15}$ & 19.8$^{+2.4}_{-2.2}$ \\
& r & 0.61$^{+0.11}_{-0.13}$ & 0.77$^{+0.06}_{-0.14}$ & 21.8$^{+0.6}_{-0.7}$ \\
& i & 0.34$^{+0.15}_{-0.09}$ & 0.73$^{+0.15}_{-0.21}$ & 19.3$^{+2.8}_{-2.4}$ \\
& z & 0.42$^{+0.17}_{-0.09}$ & 0.71$^{+0.14}_{-0.18}$ & 19.65$^{+1.6}_{-1.5}$ \\
\hline
Spot 2 & g &  0.54$^{+0.26}_{-0.17}$ & 0.67$^{+0.17}_{-0.28}$ & -52$^{+5}_{-4}$ \\
& r & 0.52$^{+0.13}_{-0.13}$ & 0.74$^{+0.09}_{-0.18}$ & -50.7$^{+1.7}_{-2.1}$ \\
& i & 0.40$^{+0.15}_{-0.10}$ & 0.61$^{+0.19}_{-0.25}$ & 
-51.4$^{+2.5}_{-2.1}$ \\
& z & 0.38$^{+0.15}_{-0.09}$ & 0.62$^{+0.19}_{-0.24}$ & 
-52.6$^{+2.2}_{-2.3}$ \\
\hline
    \end{tabular}
    \caption{Starspot parameters derived from the MCMC fit for the two detected spots. The table includes the spot radius in planetary radii ($R_{spot}/R_{plan}$), intensity relative to the stellar disk center, and longitude (in degrees) for both spots per photometric band. The uncertainties reflect the range of possible values obtained through the MCMC analysis.}
    \label{tab:spot}
\end{table}

%--------------------------------------------------------------------
\section{Temperature estimate}\label{sec:temp}

To determine the temperature of the starspots observed on CoRoT-2, we employed three different approaches, each providing independent estimates while leveraging different assumptions and models. First, we applied a blackbody fitting method, where the spot temperature was derived by fitting the observed intensities in different wavelength bands to a Planck function. Second, we utilized PHOENIX stellar atmosphere models, which offer a more realistic representation of stellar spectra by incorporating opacity effects and line blanketing, allowing for an  estimate of the spot temperature. Lastly, we implemented a simultaneous multiwavelength fitting approach, where all four photometric bands (g, r, i, and z) were fitted under the assumption of a single spot temperature, longitude, and radius across all bands, reducing degeneracies and improving robustness. The following subsections detail each of these methods, their underlying assumptions, and the resulting temperature estimates.

\subsection{Individual spot temperature fitting as blackbody for each photometric bands}

\begin{figure}
    \centering
\includegraphics[width=0.9\linewidth]{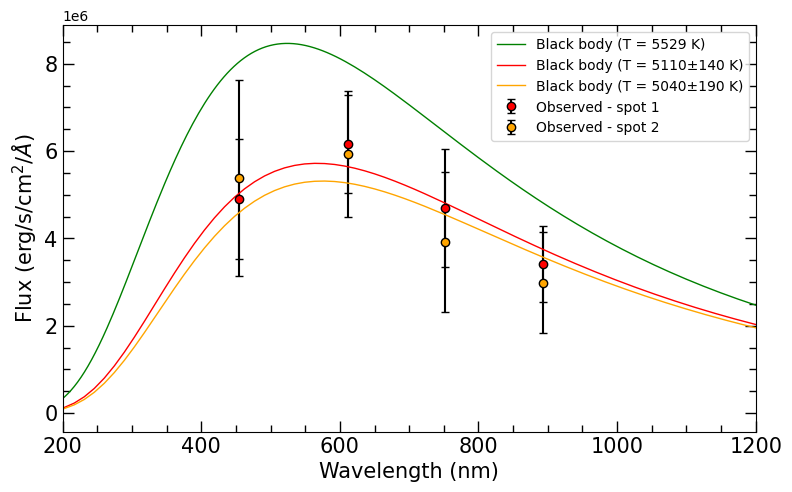}
    \caption{Starspot temperature estimation using blackbody fitting. The red and orange dots represent the observed spot intensities in the four photometric bands (g, r, i, and z), whereas the red and orange curves correspond to the best-fit blackbody spectra, yielding temperatures of $5110\pm140$ K for spot 1 (longitude of $20^\circ$) and $5040\pm190$ K for spot 2 (longitude of $-50^\circ$). The stellar photosphere is assumed to have an effective temperature of 5529~K (green curve).}
    \label{fig:bb_spot1}
\end{figure}

After calculating the intensity of the spot in each wavelength band (see Table~\ref{tab:spot}), the spot's temperature is estimated by assuming blackbody emission for both the stellar photosphere and the spot.
The intensity of the stellar photosphere, measured in erg/s/cm$^2$/\AA, was determined using an effective temperature $T_{eff} = 5529$ K \citep{Stassun2019}, whereas the spot's brightness, also in erg/s/cm$^2$/\AA, was derived based on the modeled intensity values provided in Table~\ref{tab:spot}, employing the package \texttt{Synthetic Photometry (synphot)}\footnote{\url{https://synphot.readthedocs.io/en/latest/}} \citep{2018ascl.soft11001S}. 
The spot brightness is represented by red and orange dots, with their error bars, in Figure~\ref{fig:bb_spot1} for Spots 1 and 2, respectively.
Also shown in this figure is the blackbody curve of the 5529 K star as a green curve.
These estimated brightness values were then fitted with a Planck curve using MCMC, resulting in a temperature of 5110$\pm$140 K for Spot 1, as depicted by the red curve in Figure~\ref{fig:bb_spot1}. The same modeling was performed for the second spot, with a temperature of 5040$\pm$190 K (orange curve), slightly cooler than the other spot.
The details of the estimated fluxes of both spots and the temperature estimates for the blackbody model, alongside the corresponding Bayesian Information Criteria (BIC) for each fitted model, are listed in Table~\ref{tab:bb}.

\begin{table}[]
    \centering
    \begin{tabular}{c|cccc}
\hline
\hline
Spot & Band & $\lambda_{pivot}$ & Flux & Temperature \\
& & (nm) & ($10^6$ erg/s/cm$^2$/\AA) & (K) \\
\hline
1 & g & 454.2 & 4.9$\pm{1.4}$ & $5110\pm 140$  \\
& r & 611.9 & 6.2$\pm{1.1}$ \\
& i & 751.9 & 4.7$\pm{1.3}$ \\
& z & 892.8 & 3.4$\pm{0.9}$ \\
\hline
2 & g & 454.2 & 5.4$\pm{2.2}$ & $5040\pm190$ \\
& r & 611.9 & 5.9$\pm{1.4}$ & \\
& i & 751.9 & 3.9$\pm{1.6}$ \\
& z & 892.8 & 3.0$\pm{1.2}$ \\
\hline
    \end{tabular}
    \caption{Starspot temperatures derived from the MCMC fit of a blackbody model for the two detected spots. The table includes the  wavelength ($\lambda_{pivot}$) of photometric band, spot flux, estimated spot temperature (K) and BIC of the model fit for both spots. The uncertainties reflect the range of possible values obtained through the MCMC analysis.}
    \label{tab:bb}
\end{table}

\subsection{Individual spot temperature fitting using PHOENIX spectra for each photometric band}

\begin{figure}
    \centering
\includegraphics[width=0.9\linewidth]{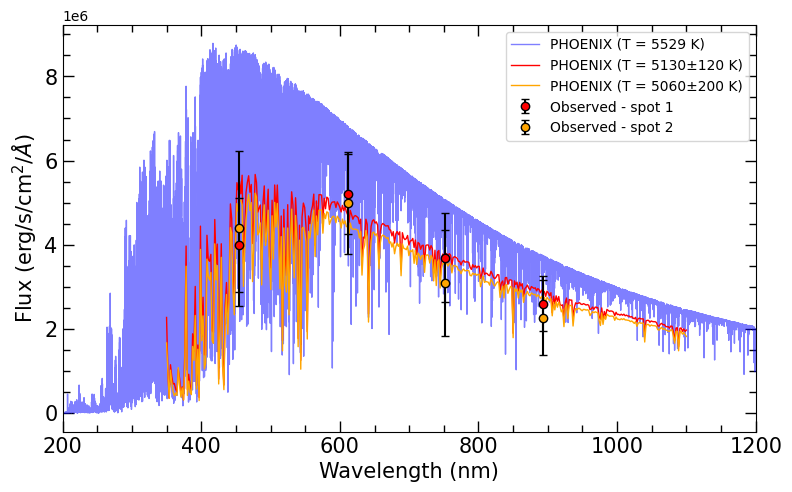}
    \caption{Starspot temperature estimation using PHOENIX stellar atmosphere models. The blue curve shows the PHOENIX spectrum for a 5529~K star (log(g) = 4.48, [Fe/H] = -0.04, [$\alpha$/M] = 0.0). The red dots represent the observed spot intensities in the four photometric bands (g, r, i, and z). The red and orange curves correspond to the best-fit PHOENIX spectra, yielding temperatures of $5130\pm120$ K for spot 1 (longitude of $20^\circ$) and $5060\pm200$ K for spot 2 (longitude of $-50^\circ$).}
    \label{fig:phoenix}
\end{figure}

The PHOENIX spectra can serve as a more accurate reference for the star \citep{Husser2013}, replacing the blackbody approximation. The \texttt{Phoenix Models for Synphot} provides stellar spectra for different stellar metallicities ([Fe/H]), which are available on their official website\footnote{\url{https://www.stsci.edu/hst/instrumentation/reference-data-for-calibration-and-tools/astronomical-catalogs/phoenix-models-available-in-synphot}} \citep{2013ascl.soft03023S}. For this analysis, we selected the spectrum of a 5529~K (log(g) = 4.48, [Fe/H] = -0.04, [$\alpha$/M] = 0.0) star, shown in blue on Fig.~\ref{fig:phoenix}, employing the package \texttt{Synthetic Photometry for HST (stsynphot)} \footnote{\url{https://stsynphot.readthedocs.io/en/latest/}} \citep{2020ascl.soft10003S}. 
We estimated the starspots fluxes by scaling the PHOENIX spectrum of a 5529 K star at the central wavelengths of the g, r, i, and z photometric bands used in the observations (red and orange dots with error bars, corresponding to Spot 1 and 2, respectively, overplotted on the PHOENIX spectra on Figure~\ref{fig:phoenix}) using the spot intensities listed in Table~\ref{tab:spot}.
These estimated brightness values of each spot were then fitted with a PHOENIX model (red and orange curves on Figure~\ref{fig:phoenix}), resulting in a temperature estimate of 5130$\pm$120 K for Spot 1, and 5060$\pm$200 K for Spot 2.
The details are listed, alongside the corresponding Bayesian Information Criteria (BIC) for each fitted model, in Table~\ref{tab:phoenix}. As can be seen, these values are very close to those obtained by the black body model (see Table~\ref{tab:bb}).

\begin{table}[]
    \centering
    \begin{tabular}{c|cccc}
\hline
\hline
Spot & Band & $\lambda_{pivot}$ & Flux & Temperature \\
& & (nm) & ($10^6$ erg/s/cm$^2$/\AA) & (K) \\
\hline
1 & g & 454.2 & 4.0$\pm{1.1}$ & $5130\pm120$ \\
& r & 611.9 & 5.2$\pm{0.9}$ \\
& i & 751.9 & 3.7$\pm{1.1}$ \\
& z & 892.8 & 2.6$\pm{0.7}$ \\
\hline
2 & g & 454.2 & 4.4$\pm{1.8}$ & $5060\pm200$  \\
& r & 611.9 & 5.0$\pm{1.2}$ & \\
& i & 751.9 & 3.1$\pm{1.3}$ \\
& z & 892.8 & 2.3$\pm{0.9}$ \\
\hline
    \end{tabular}
    \caption{Starspot temperatures derived from the MCMC fit of a PHOENIX atmospheric model for the two detected spots. The table includes the  wavelength ($\lambda_{pivot}$) of photometric band, spot flux, estimated spot temperature (K) and BIC of the model fit for both spots. The uncertainties reflect the range of possible values obtained through the MCMC analysis.}
    \label{tab:phoenix}
\end{table}

\subsection{Assuming the same spot's temperature, radius, and longitude for all photometric band}\label{sec:multi}

Next, we fit all 4 photometric bands, g, r, i, and z, simultaneously for a spot with  the same temperature, longitude, and radius across all 4 bands. For this, we considered blackbody emission for the spot which intensity, $I_{spot}$, is related to the spot ($T_{spot}$) and stellar ($T_{eff}$) temperatures by \citep{zaleski2022dynamo}:
\begin{equation}
     \frac{I_{spot}}{I_{phot}} = \frac{\exp\left(\frac{hc}{{\lambda}K_{B}T_{\rm eff}}\right) - 1}{\exp\left(\frac{hc}{{\lambda}K_{B}T_{\rm spot}}\right) - 1},
     \label{eq:intratio}
\end{equation}
\noindent where $K_B$ and $h$ are Boltzmann and Planck constants, respectively, $\lambda$ is the relevant wavelength, $I_{phot}$ is  the central stellar intensity,  and $T_{eff}$ is the effective temperature of the star (5529 K). Thus, for each temperature guess of MCMC, the spot intensities were estimated accordingly and used in the fit of \texttt{ECLIPSE}. The priors were the same as before, except that instead of priors for the intensity, we restricted the temperature prior to be between 3000 K and 5500 K.

The results of the MCMC fit are shown in Figure~\ref{fig:spot_temp} for both spots. 
As can be seen from the figure, the best fit yields 
radius for each spot of 0.65 $R_{Plan}$ and 0.58 $R_{Plan}$, and a temperature of $5280^{+80}_{-130}$ K for  Spot 1 at longitude $21^\circ$, whereas a slightly cooler temperature of $5220^{+110}_{-230}$ K was obtained for Spot 2 at longitude  $-52^\circ$. 

Table~\ref{tab:all_temp} resumes all the temperature estimates, including the three methods, for both spots. As can be seen, despite Spot 2 being slightly cooler than Spot 1, the temperature of both  spots is similar, within the error bars, independent of the model. On the other hand, the multiwavelength method yields larger spots with temperatures about 100 K warmer than the other two models, which agree well with each other.

\begin{table}[]
    \centering
    \begin{tabular}{c|ccc}
\hline
\hline
 Spot  & $T_{Blackbody}$ (K) & $T_{PHOENIX}$ (K) & $T_{multi\lambda}$ (K) \\
 \hline
1 & $5110\pm140$ & $5130\pm120$ & $5280^{+180}_{-130}$ \\
2 & $5040\pm190$ & $5060\pm200$ & $5220^{+110}_{-230}$ \\
\hline
    \end{tabular}
    \caption{Estimated temperature of the two spots by the blackbody, PHOENIX spectral, and multiwavelength fittings with MCMC.}
    \label{tab:all_temp}
\end{table}

%--------------------------------------------------------------------
\section{Discussion and conclusions}\label{concl}

Here, we detected the signature of two spots on one transit of CoRoT-2 observed on the 20th of June of 2024 with the SPARC4 instrument on the 1.6-m telescope of OPD, using light curves obtained  simultaneously in the g, r, i, and z bands. These data were applied to estimate the temperature and sizes of these spots. 
By applying three different methods - blackbody, PHOENIX spectra, and a simultaneous multiwavelength fit - we obtained consistent temperature estimates (see Table~\ref{tab:all_temp}).

First, we modeled the two spots using the \texttt{ECLIPSE} code, obtaining the spots' parameters, which are listed in Table~\ref{tab:spot}. The spots' intensity (0.61 - 0.77) estimated here is higher than those obtained previously by \cite{silva2010}. 
Once the intensity of the spots was determined, the spot brightness was estimated  assuming the stellar photosphere to be a blackbody of $T_{eff} = 5529$ K. Then these brightness values in the four photometric bands were fit by a blackbody (Figure~\ref{fig:bb_spot1}), yielding temperatures of 5110 $\pm$ 140 K and 5040 $\pm$ 190 K for the two spots at longitudes $\sim20^\circ$ and $\sim-50^\circ$, respectively.

The PHOENIX spectral method provided similar values for the spots' temperatures, of 5130 $\pm$ 120 K and 5060 $\pm$ 200 K for the two spots, with the inclusion of stellar opacity effects. Lastly, the simultaneous multiwavelength fitting method (with the same radius, longitude, and temperature for each spot) resulted in temperatures of $5280^{+80}_{-130}$ K and $5220^{+110}_{-230}$
K. 
Irrespective of the method, these temperatures agree with each other and are higher than the average temperature of $4700 \pm 300$ K, but within the range of 4340 and  5260 K found for the 369 spots by \cite{silva2010}.

In a sunspot, the umbra (the darkest central region) has an average temperature of around 3700–4000 K, while the penumbra (the lighter, surrounding area) is slightly warmer, with temperatures averaging between 5000–5500 K \citep{Solanki2003}. These temperatures are lower than the surrounding solar photosphere, which is about 5778 K. The difference in temperature between the umbra and penumbra is due to variations in magnetic field strength and structure, which affect the energy transport within these regions. 

In the model adopted here, the lack of spatial resolution leads to the assumption of isothermal starspots. 
In our analysis of CoRoT-2, we derive spot temperatures between 5040 K and 5280 K for an effective stellar temperature of 5529 K, corresponding to $\Delta T$ in the range of -489 K to -249 K. These values are compatible with temperature contrasts observed on the active M dwarf ($\Delta T$ = -480 to -280 K) by \cite{Libby-Roberts2023}, as well as those expected from solar penumbrae ($\Delta T \simeq -400$ to -250 K; \citealt{Solanki2003}). This further reinforces the physical plausibility of our fitted spot temperatures and supports the interpretation that CoRoT-2 exhibits penumbra contrasts similar to solar-like spot.

In terms of size, individually fitting the transit light curves for different wavelengths resulted in the spots having radii ranging from 0.34 to 0.61 planetary radii, or 38 - 69 Mm, which are significantly larger than typical sunspots (5 - 20 Mm, \cite{Solanki2003}), but in agreement with sizes  previously found for CoRoT-2 by \cite{silva2010}. Additionally, when all the wavelengths light curve residuals were fit simultaneously, the two spots' radii were $0.66^{+0.15}_{-0.16}$ $R_{plan}$ and $0.57 \pm 0.13$ $R_{plan}$, values  still within the range of 0.2 to 0.7 $R_{plan}$ found by \cite{silva2010}.

In modeling starspots during planetary transits using the \texttt{ECLIPSE} model, a known degeneracy exists between the spot radius and intensity parameters: a small, colder spot can produce the same flux deficit as a larger, warmer spot. This degeneracy arises because the overall effect on the transit light curve depends on the combined effect of the spot's size and brightness contrast relative to the stellar photosphere. As a result, it is challenging to uniquely determine both the spot's temperature and size simultaneously from transit data alone, leading to potential inaccuracies in the estimated spot temperature. 

Nevertheless, these results confirm that CoRoT-2 exhibits strong magnetic activity, with spots that are relatively warm and extensive in size, consistent with previous studies on this highly active young star. 
In summary, the methodology presented here demonstrates the potential of multiwavelength photometry for improving starspot characterization, which is crucial for refining stellar activity models and exoplanet transit corrections. Further observations of this kind, covering a broader sample of active stars, are necessary to better constrain the temperature distribution of starspots and their impact on stellar and exoplanetary studies.
CoRoT-2 is scheduled to be observed by the ARIEL mission, which will provide new opportunities for the characterization of its planetary system.

\appendix 

\section{Corner plots of the spot parameters' fit}

\begin{figure}[h]
    \centering
\includegraphics[width=0.48\linewidth]{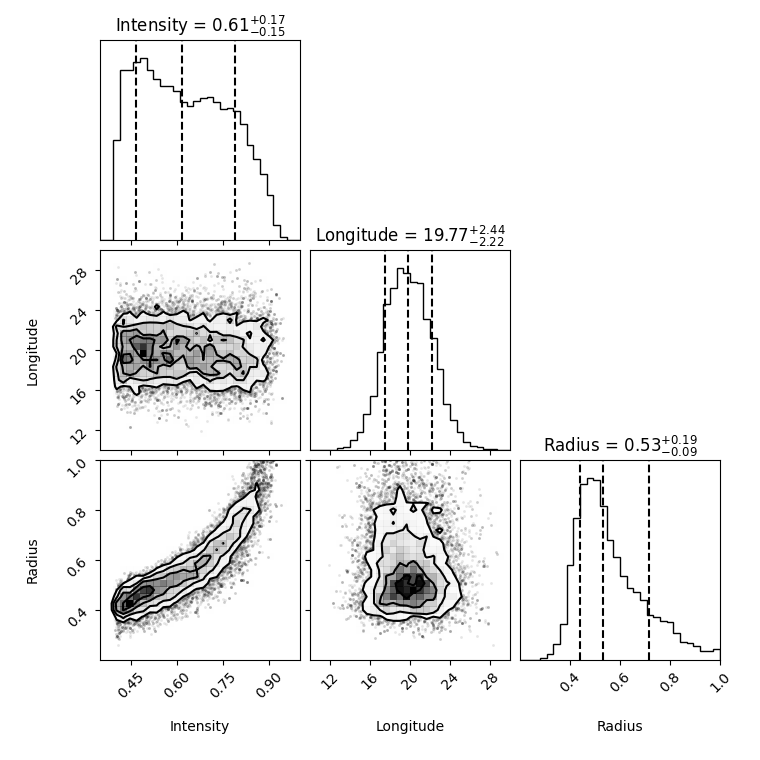}
\includegraphics[width=0.48\linewidth]{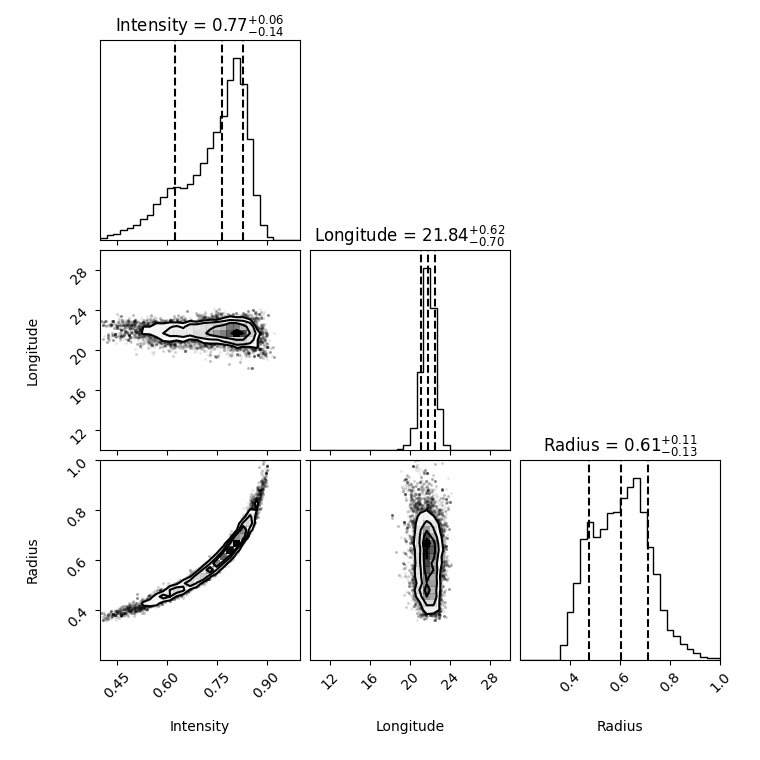}
\includegraphics[width=0.48\linewidth]{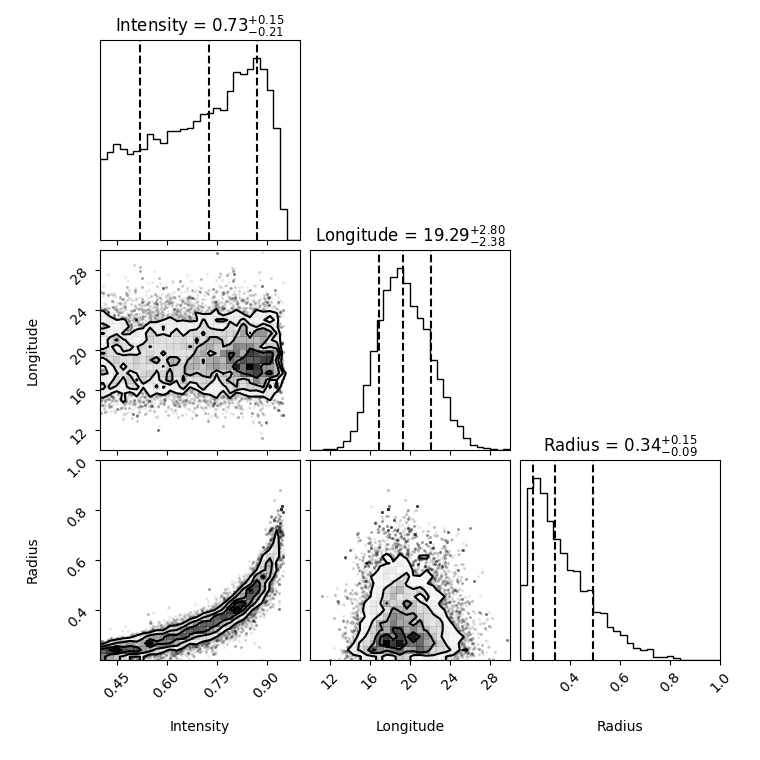}
\includegraphics[width=0.48\linewidth]{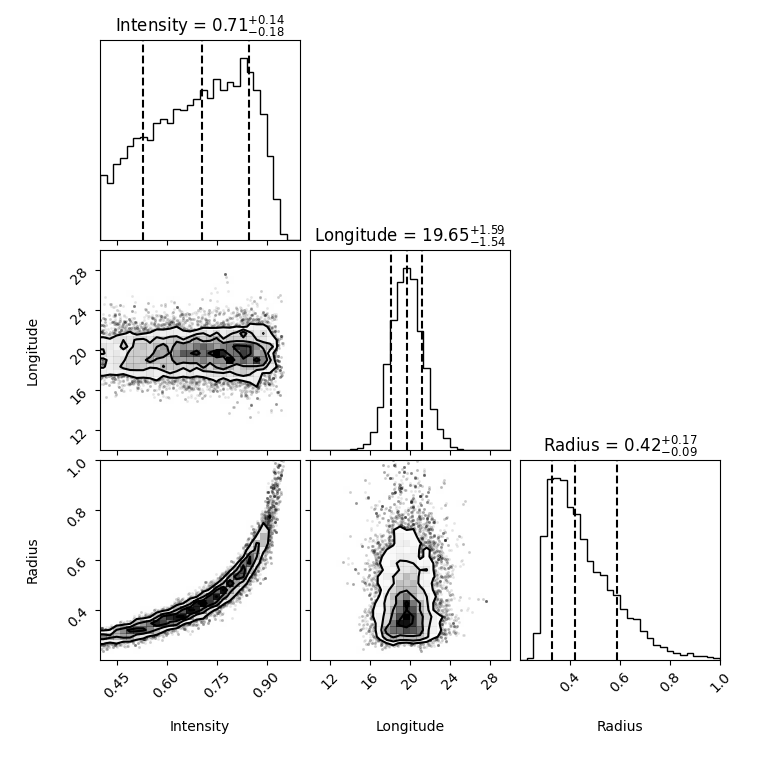}
    \caption{Results of the MCMC fit to spot at longitude $20^\circ$ in the residuals of the transit light curve for the four photometric bands. The fit parameters of the spot are: radius (in units of planetary radius), intensity, and longitude. The plots show the MCMC samples and the respective posterior distributions of each free parameter.}
    \label{fig:spot1}
\end{figure}

\begin{figure}[b]
    \centering
\includegraphics[width=0.48\linewidth]{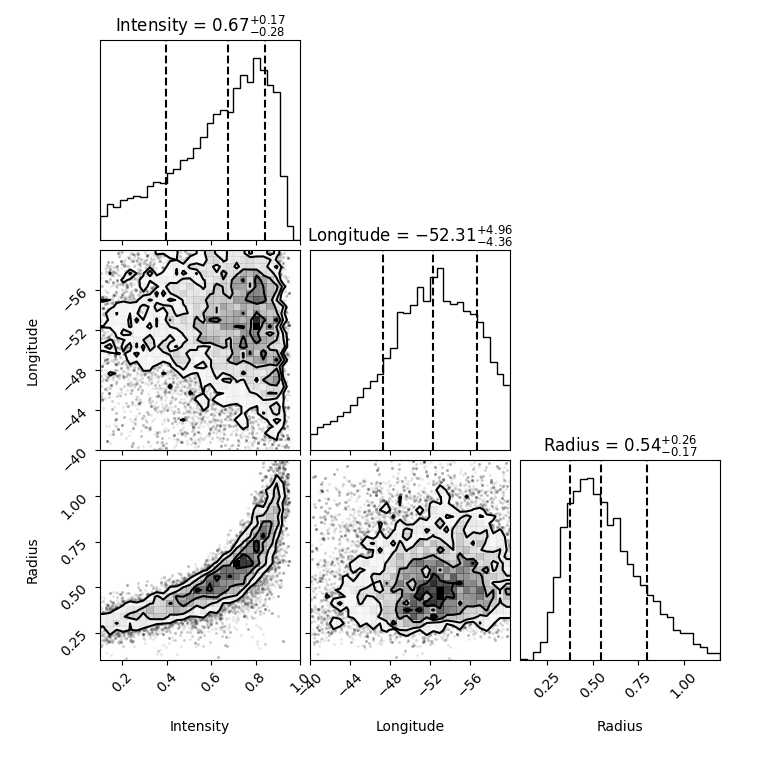}
\includegraphics[width=0.48\linewidth]{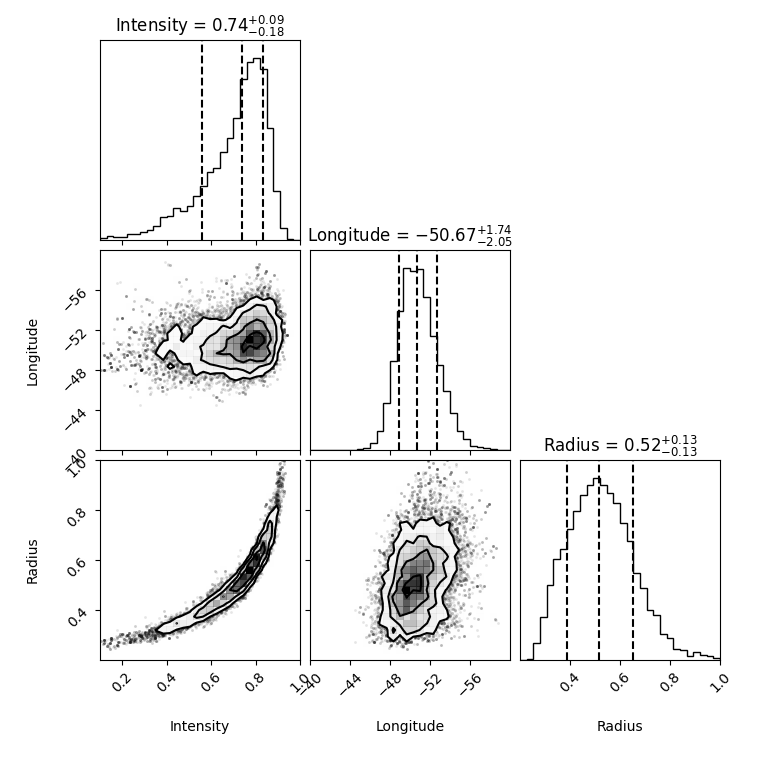}
\includegraphics[width=0.48\linewidth]{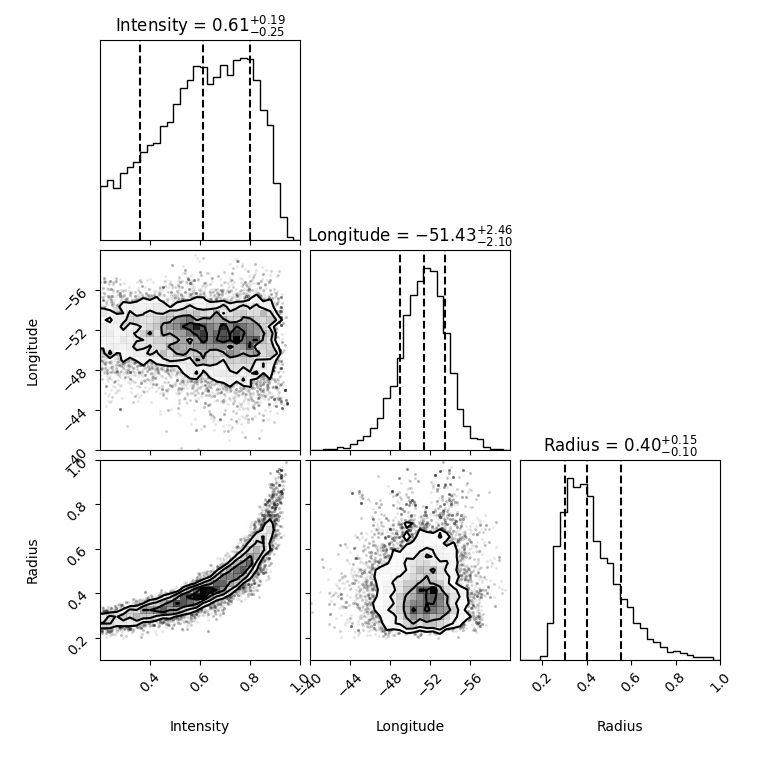}
\includegraphics[width=0.48\linewidth]{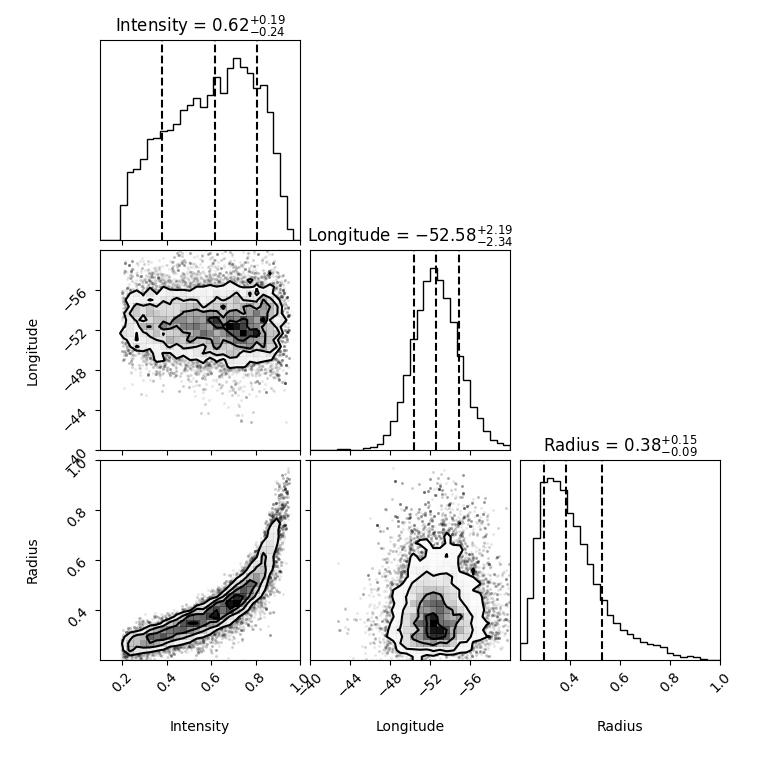}
    \caption{Same as Fig.~\ref{fig:spot1} for the second spot at longitude $-50^\circ$.}
    \label{fig:spot2}
\end{figure}

Figures~\ref{fig:spot1} and \ref{fig:spot2} display the corner plots resulting from the MCMC fitting of the spot parameters for each detected starspot in the transit light curve. These plots show the sample distributions and posterior probabilities for each of the spot’s free parameters namely, radius (in units of planetary radius), intensity (relative to the stellar disk center), and longitude. 

\section{Corner plot of the multiwavelength temperature fit}

\begin{figure}[h]
    \centering
    \includegraphics[width=0.48\linewidth]{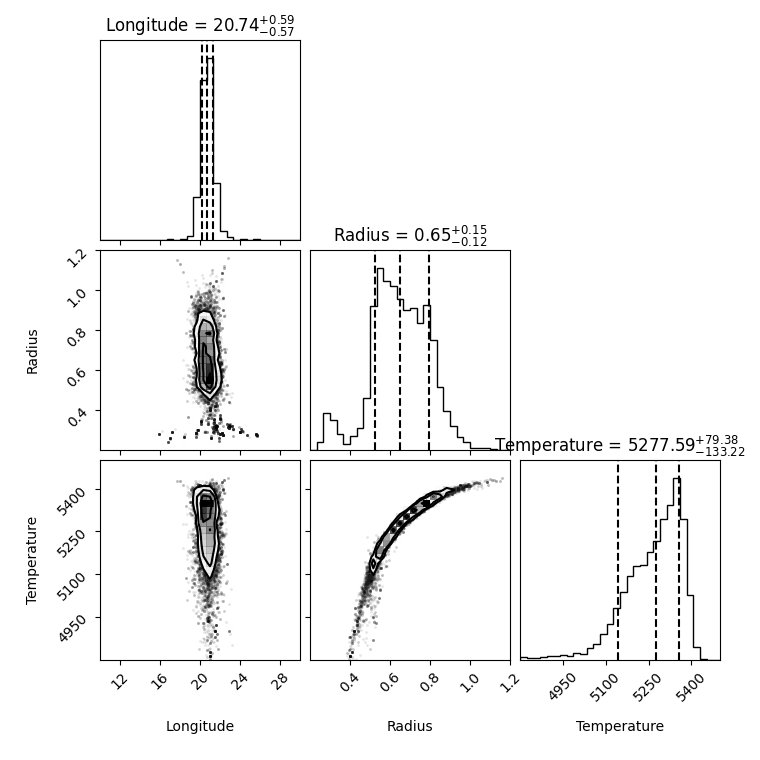}
    \includegraphics[width=0.48\linewidth]{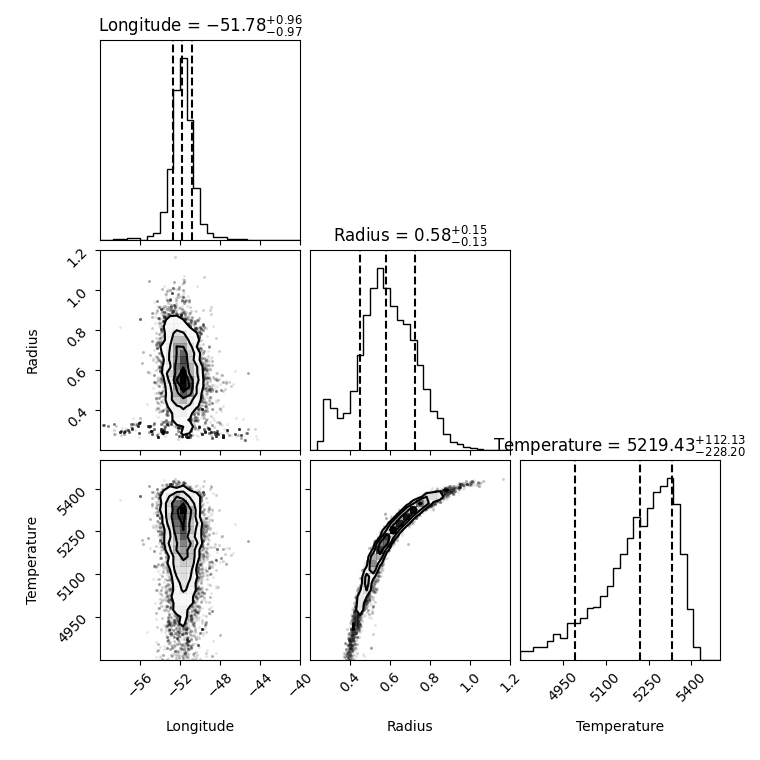}
    \caption{MCMC results of the simultaneous multiwavelength fitting for spot temperature estimation. The best-fit models for both spots assume a single temperature, longitude, and radius across all four photometric bands. The left panel shows the fit for the spot at longitude $20^\circ$, while the right panel shows the fit for the spot at longitude $-50^\circ$. }
    \label{fig:spot_temp}
\end{figure}

Figure~\ref{fig:spot_temp} presents the corner plots from the simultaneous multiwavelength MCMC fitting, in which a single temperature, radius, and longitude are assumed for each spot across all four photometric bands.
These plots summarize the joint posterior distributions of the spot parameters, providing the spot temperature estimates derived from fitting all available bands at once. The figure shows the best-fit models for the two spots identified at longitudes $\sim20^\circ$  and $\sim-50^\circ$, in the left and right panels respectively.

\begin{acknowledgements}
This paper uses data obtained with the Simultaneous Polarimeter and Rapid Camera in 4 bands (SPARC4), installed on the 1.6-m telescope at the Observatório do Pico dos Dias (OPD), managed by the Laboratório Nacional de Astrofísica (LNA) under the Ministério da Ciência, Tecnologia e Inovação (Brazil). 
A.O.K. acknowledges funding from CAPES. This study was financed in part by the Coordenação de Aperfeiçoamento de Pessoal de Nível Superior - Brasil (CAPES) - Finance Code 001.
A.V. acknowledges partial funding from FAPESP (2021/02120-0).
E.M. acknowledges funding from FAPEMIG under project number APQ-02493-22 and a research productivity grant number 309829/2022-4 awarded by the CNPq.
C.V.R. thanks the Brazilian Space Agency (AEB) by the support from PO 20VB.0009 and the Brazilian National Council for Scientific and Technological Development – CNPq (Proc:	310930/2021-9).

\end{acknowledgements}

%\bibliography{Biblio_spot}
%\bibliographystyle{aasjournal}

\end{document}